# Spectroscopic characterization of graphene films grown on Pt (111) surface by chemical vapor deposition of ethylene

E. Cazzanelli [a,b,*], T. Caruso [a,b], M. Castriota [a,b], A. R. Marino [a], A. Politano [a], G. Chiarello [a,b], M. Giarola [c] and G. Mariotto [c]

[a] *Dipartimento di Fisica, Università della Calabria, Ponte Bucci, Cubo 31C, 87036 Arcavacata di Rende (CS), Italy*
[b] *Consiglio Nazionale Interuniversitario di Scienze Fisiche della Materia, Via della Vasca Navale, 84, 00146 Roma, Italy*
[c] *Dipartimento di Informatica, Università di Verona, Strada le Grazie 15 – Ca' Vignal 2, 37134 Verona (VR), Italy*

**Abstract**

This work reports the peculiar properties of a graphene film prepared by the chemical vapor deposition (CVD) of ethylene in high vacuum on a well oriented and carefully cleaned Pt(111) crystal surface maintained at high temperature. In-situ and ex-situ characterization techniques (low energy electron diffraction, high resolution electron energy loss spectroscopy, scanning electron microscopy and Raman micro-spectroscopy) used here indicate the prevalence of single layer regions and the presence of two different orientations of the graphene sheets with respect to the Pt(111) substrate. In most of the deposited area evidence is found of a compressive stress for the graphene lattice, as a net result of the growth process on a metal substrate. This graphene film grown on Pt(111) exhibits a lower degree of order and of homogeneity with respect to the exfoliated graphene on Si/SiO$_2$, as it is found generally for graphene on metals, but several characterization techniques indicates a better quality than in previous deposition experiments on the same metal substrate.

* Corresponding author
E-mail address: enzo.cazzanelli@fis.unical.it (E. Cazzanelli)

1. Introduction

The epitaxial growth of large, highly ordered graphene monolayers is a prerequisite for most technological applications of this novel material [1]. In particular, the growth of graphene on metal substrate is a very promising route for the synthesis of graphene samples with high crystalline quality [2-6]. Graphene/metal interfaces are usually divided into two classes with respect to the interaction between the graphene sheet and the metal substrate. The first class is characterized by "strong" interaction, as for graphene grown on Ru(0001) [7, 8], Ni(111) [9] and Re(0001) [10]; for such substrates, only a single in-plane orientation of the graphene layer has been found [8]. In contrast, different in-plane graphene orientations occur in the other class of interfaces, showing a "weak" interaction: graphene domains assume several azimuthal orientations on substrates such as Pt(111) [11-15], Ir(111) [16], and Pd(111) [17]. Therefore, the study of these possible in-plane orientations is a benchmark for graphene/metal interface characterization.

Recently, it has been reported [18] that two different in-plane orientations of graphene on the same Ir(111) substrate exhibit striking dissimilarities in their electronic properties: *i.e* graphene domains tilted by 30° with respect to the Ir substrate (R30) are more weakly bonded to the metal with respect to domains aligned with Ir lattice (R0). Angle-Resolved Photo Electron Spectroscopy (ARPES) reveals that, for R30 domains, the Ir substrate does not significantly modify $\pi$ bands of the free-standing graphene. On the contrary, R0 domains exhibit an appreciable band gap. Moreover, R0 domains do not show Raman-active modes, because of the quenching of the resonance conditions, due to the hybridization of the $\pi$ bands with metal *d*-states; on the contrary, in R30 graphene Raman-active phonons are observed [18]. Raman spectroscopy provides important information on these systems. In particular, single layer graphene exhibits typical spectral patterns, which allow to discriminate it from multilayer graphene [19, 20, 21]. Previous studies [18, 20-26] also reveal how the spectral features of the relevant Raman bands of graphene, like frequency, intensity and shape, are affected by the interaction with the substrate: for instance, in graphene flakes deposited on Si/SiO$_2$ substrates by means of mechanical exfoliation [19, 21, 27], the peak-frequency of G band is close to the typical value of graphite, appreciably lower than in graphene layers obtained from SiC [28, 29] or deposited by CVD on metal surfaces [18, 23-24, 30]. Moreover, the Raman cross section depends on the substrate nature, as well as on the relative orientation between the basal plane axes of carbon sheet and the substrate surface [18]. Raman spectra have been previously collected for graphene deposited on platinum thin films [31] and on a multifaceted surface [23] with several crystallographic orientations. For well oriented Pt (111) crystalline substrate, a micro-Raman mapping of graphene grown by surface segregation of carbon-doped platinum is reported [32], but the signal was very weak, making difficult to ascertain the homogeneity and continuity of that sample.

Herein, we present combined results of several investigation techniques on a film of graphene grown by CVD over a Pt (111) surface: in-situ, low energy electron diffraction (LEED), high-resolution electron energy loss spectroscopy (HREELS) and, ex-situ, scanning electron microscopy (SEM) and micro-Raman spectroscopy.

2. **Experimental**

Graphene sheets were obtained by dosing ethylene onto the clean Pt(111) substrate at 1150 K in a ultra-high vacuum (UHV) chamber, where the film was also characterized by some surface techniques. The substrate was a single crystal of Pt(111), delivered from MaTecK GmbH (Germany). It was cleaned by repeated cycles of ion sputtering and annealing at 1300 K. Surface cleanliness and order were checked using Auger Electron Spectroscopy (AES) and electron diffraction (LEED) measurements, respectively.

HREELS experiments were performed by using an electron energy loss spectrometer (Delta 0.5, SPECS). The energy resolution of the spectrometer was degraded to 5 meV so as to increase the signal-to-noise ratio of loss peaks.

SEM analysis was performed by a FEI QUANTA FEG 400 F7 microscope.

Raman spectroscopy measurements were carried out in air by using different equipments, with different excitation wavelengths, from red (647.1 nm) to blue (488 nm). The most significant data were obtained by using a confocal microprobe apparatus: an Olympus microscope interfaced to a triple monochromator (Horiba-Jobin Yvon, model T64000), mounting holographic gratings having 1800 lines/mm, set in double-subtractive/single configuration, and equipped with a CCD (256x1024 pixels) detector, cooled by liquid nitrogen. A high magnification objective 100X was used to focus the laser beam onto the sample surface, to maximize the spatial resolution and the signal gain. Moreover, a confocal approach has been adopted to reduce the background scattering with respect to graphene signal. Polarized micro-Raman spectra were collected at room temperature from 180° scattering geometry, mostly in crossed XY polarization; the maximum output laser power, for the 488 nm and 514.5 nm lines, was 20 mw, and lower laser powers were used for the other wavelengths. To get a satisfactory signal-to-noise ratio long integration times were necessary, typically 900 s, for the thinnest sample regions, corresponding to one or few graphene layers.

3. Results and Discussion

3.1 **In-situ measurements**

The graphene growth on Pt(111) substrate was monitored in-situ by LEED spectroscopy. This analysis suggests that the saturation of a mono layer graphene (MLG) on the Pt (111) substrate was reached upon an exposure of ethylene of $3 \cdot 10^{-8}$ mbar for ten minutes (24 Langmuir). As demonstrated also by in-situ low-energy electron microscopy studies [33], no nucleation and growth of additional graphene sheets beyond the MLG occurs on Pt (111) [12, 14, 31, 33-37].

The presence of well-resolved spots in the LEED pattern (see Figure 1) is a clear fingerprint of the order of the MLG over-structure. The attained LEED pattern is essentially similar to that reported by Gao et al. [12]. The ring pattern indicates the existence of different domains. Nonetheless, preferred orientations aligned with the substrate (R0) are clearly distinguished. Despite the presence of other domains, the predominance of R0 domains has been clearly inferred by the analysis of phonon dispersion measurements performed along specific directions of the sample, which can correspond to the $\bar{\Gamma}-\bar{K}$ or to the $\bar{\Gamma}-\bar{M}$ directions of graphene reciprocal lattice, depending on the orientation of the explored domain, R30 or R0, respectively. In fact, the best fit of experimental data points on the theoretical dispersion curves was obtained by assuming a dominant R0 orientation. The characterization of the MLG was carried out by measuring phonon modes which are a fingerprint of graphene formation [37-39], as shown in Figure 2. The occurrence of well-resolved ZA (out-of-plane acoustic), ZO (out-of-plane optical), LA (longitudinal acoustic), LO (longitudinal optical) and TO (transverse optical) phonons ensures of the good order and crystalline quality of the graphene sheet.

The analysis of both the LEED patterns and phonon modes dispersion suggests a weak interaction between MLG and the underlying Pt substrate, in fair agreement with previous works [12, 14, 33, 40]. Accordingly, MLG can be considered as a quasi-freestanding sheet physisorbed on the Pt substrate.

**3.2 Ex-situ investigations**

The as-deposited graphene layer was later characterized ex-situ by SEM and micro-Raman spectroscopy. The surface morphology appears to SEM exploration homogeneous across all the sample (1x1 cm$^2$). SEM images (see Figure 3) show a full coverage of the substrate surface by the graphene which forms a network of wrinkles (darker horizontal and vertical lines in Figure 3). The wrinkles network develops on a micrometric length scale. Its morphology matches that obtained by low-energy electron microscopy (LEEM) measurements for graphene grown on Pt (111) by carbon segregation from the Pt (111) substrate and other metallic substrates [33]. In addition, zones of average micrometric size, showing two different shades, are observed on the surface, probably due to different graphene domains orientations. Their presence could also be correlated to wrinkles network. Another observed detail concerns the growth of randomly distributed sub-micrometric islands, suggesting the formation of thicker graphitic structures.

A systematic micro-Raman spectroscopic characterization has been carried out on several sample regions. Preliminary measurements were performed with red excitation light (633 nm) by using a micro-Raman set-up Horiba-Jobin Yvon (model LabRam), without detecting any significant signal. Better results were obtained by using a T-64000 triple-monochromator, with a Kr/Ar ion laser source providing several excitation lines through the visible region. Analysis of spectra excited by different laser lines (not shown here) indicate a remarkable dependence of Raman intensity on the excitation energy: Raman spectrum of graphene was clearly observable by using blue (488.0 nm) and green (514.5 nm) laser lines, a drop of intensity was observed for yellow line (568.2 nm), and a barely detectable signal was obtained for a red line (647.1 nm). Taking into account the difficulty to compare quantitatively the Raman cross sections of different samples, a rough estimation of the order of magnitude can be made about the intensity change with respect to previous measurements on exfoliated graphene on Si/SiO$_2$, by using the intensity ratio with Raman-active modes of air molecules; in fact Raman peaks due to O$_2$ and N$_2$ stretching, occurring at 1556 cm$^{-1}$ and 2331cm$^{-1}$, respectively [41], originate from the air volume above the graphene sheet within the laser waist size, and also provides useful frequency standards. The Raman intensity of graphene on Pt (111) turns out between one and two orders of magnitude weaker than the one of exfoliated graphene on Si/SiO$_2$. Similar intensity decreases, with respect to Si/SiO$_2$ substrate, are reported for graphene on sputtered thin Pt film [31] and on multifaceted textured Pt foils [23].

It is well known [23, 31, 42] that a much stronger Raman signal can be obtained by transferring a graphene film grown on some metal to a new Si/SiO$_2$ substrate, following a methodic recently developed [43]. However, this operation provides information about a new system, physically different from the original one, as grown on metal. If the investigation is aimed to explore the graphene-metal interaction, the weakness of the Raman signal, and even its absence constitutes a relevant information, which is lost in the transfer process. For such reason we limited our micro-Raman analysis to the as-grown graphene on Pt (111).

As for the peak frequencies, the values of D band and its overtone 2D depend remarkably on the excitation energy, while the G band position is insensitive to it, changing its frequency only because of charge doping or mechanical stresses [21]; for this reason it constitutes a good indicator of different kinds of graphene-substrate interaction. In our present investigation the explored regions of the film give frequency values of the G band in a range 1600-1605 cm$^{-1}$ (see Figure 4). In the same Figure 4 are reported as typical examples the spectra collected along a line, step by step of about 10 microns, under excitation of the 488 nm laser line in crossed XY polarization, showing the wavenumber ranges containing G band and 2D overtone. Both the mono-modal spectral shape of the 2D overtone and the comparable intensity of 2D overtone and G band support the hypothesis of a monolayer graphene for most of the explored regions, following commonly established criteria [19, 21, 44] (taking into account that in Figure 4 the G band intensity is relatively enhanced with respect to the 2D overtone because of the crossed polarization setting). The dispersion of the G band peak frequency, in this spectra sequence, is quite small (few wave-numbers), comparable with the instrumental resolution. The intensity of D band, at about 1360 cm$^{-1}$, which is a characteristic marker of structural disorder, is very low; in many spectra it does not merges out from the noise. The line-width of G band appears also to be quite homogeneous through the various regions, varying between 12 cm$^{-1}$ and 17 cm$^{-1}$ for most of the measured spots. It may be interesting to compare this range of bandwidths with the minimum values found in exfoliated graphene, down to 6-7 cm$^{-1}$ [19, 27]: it is reasonable to explain this difference in bandwidth by assuming a distribution of uniaxial strains on the graphene layers grown on Pt(111), in addition to an isotropic strain, responsible for the G band frequency up-shift from ~1585 to 1600-1605 cm$^{-1}$. This particular strain distribution can be ascribed to the high temperature processes leading to the carbon segregation and to the structural rearrangements occurring during the cooling down to room temperature, for both the graphene overlayer and the metal substrate, driven by two different thermal expansion coefficients [23, 45].

The overtone 2D band presents high values of peak frequency, about 2720 cm$^{-1}$, and a variation of the line-width within the range 32-40 cm$^{-1}$. The spectra exhibiting the narrower 2D overtone also show minimum width of the G band; therefore they can be attributed to more ordered and homogeneous regions of monolayer graphene. In fact, the strong increase of 2D bandwidth has been associated with bilayer or multilayer arrangements, even for the case of random (non Bernal) stacking [21], not showing the usual multimodal character of 2D band. The bandwidth value of about 35 cm$^{-1}$ found for some spot of graphene on Pt (111) does not exceed very much the value of 28 cm$^{-1}$ measured for monolayer exfoliated graphene [27]. It is interesting to compare such findings with others available Raman data on graphene over Pt: in the case of graphene on sputtered platinum thin film [31], the Raman bands appear very weak, as it results from intensity comparison with $N_2$ and $O_2$ stretching modes of the air; the frequencies of G and 2D bands are up-shifted at about the same values found here, but the bandwidths of the Raman modes are greater (G mode always broader than 22 cm$^{-1}$, 2D mode always broader than 40 cm$^{-1}$), indicating a higher dispersion of the stress values. In the case of graphene grown on multifaceted Pt foils (111-, 110- and 100-oriented) no frequency up-shift is observed with respect to Si/SiO$_2$ [23]; a possible explanation is the prevalence in the analyzed region of different crystal surfaces, *i.e* (100) and (110), which do not induce on the graphene layer the same strain as the Pt (111) surface during the growth.

As a first conclusion, our graphene layer grown on Pt(111) surface appears as an extended, ordered and more homogeneously strained layer, with respect to previous similar preparations [31, 32].

Moreover, by assuming a prevalence of R0 domains, as suggested by the phonon dispersion HREELS investigation, or even a comparable occurrence of R0 and R30, consistent with LEED measurements, the appreciable Raman intensity found throughout the surface of our sample indicates that both the domain orientations generate a comparable Raman signal.

Finally, the varying parameters of the Raman bands among the many explored spots seem to indicate some additional complexity. In some case, a few spots on the surface generate spectra where a distinct G band can be observed, with lower frequency values, about 1585 cm$^{-1}$. In some other case a remarkable asymmetric shape of G band still indicates the existence of regions with lower frequency G band, for a smaller fraction.

In Figure 5 typical Raman spectra are shown, for sake of comparison, collected from zones of graphene with different amounts of defects, starting from the most homogeneous region (spectrum 5a), through an intermediate case (spectrum 5b) up to the spectrum 5c, where a remarkable amount of carbon, located within the laser irradiated area, generates a second G band with lower peak frequency values. Possible sources of such down-shifted G band can be zones of graphene monolayer with an unstrained configuration, like wrinkles, reported for Pt substrate as well as for other metals [33, 45]. Another realistic hypothesis involves regions of multilayer growth, approaching locally the bulk graphite morphology, also reported on Pt(111) [32]. If the hypothesis of wrinkles is accepted, in absence of direct measurement of their typical size, this one must be supposed much smaller than that of the irradiated area, in such a way to make possible a continuous change of the Raman spectral shape as effect of the varying concentration of defective structures. On the other hand, the hypothesis of multilayered structures is supported, in our deposited film, by SEM evidences of many sub-micrometric islands, which can be associated to multilayered towers or pyramids, in contrast with previous LEEM investigations [32], which report corresponding structures of greater, micrometric size; the bimodal shape or the lower frequency asymmetry, corresponding to spectra 5c and 5b, respectively, can be justified by assuming such small sizes of the multilayered islands, consistently with the SEM analysis but escaping the direct optical microscopy exploration.

## 4. Summary

A detailed investigation carried out by using several techniques has been performed on graphene deposited on a Pt (111) substrate via a particular CVD method, described above, that, in principle, can provide different structural order and domains distribution with respect to other similar systems.

In the sample here analyzed, most of the surface of the Pt (111) substrate appears well coated with monolayer graphene, which is more ordered and homogeneous than for similar preparations previously reported. A compressive strain is found for the graphene as the effect of the growth process, and a widespread occurrence of wrinkles on the surface is observed. Two possible orientations are found for the graphene overlayer: a well defined and homogeneous R0, with the same orientation of the substrate, and another tilted one, R30, with the maximum of the distribution of tilt angles at 30°, but with an appreciable spread around that value. Both these domain orientation seem to generate a comparable Raman signal. In some regions the occurrence of small, sub-micrometric, multilayered pyramids can be postulated, to explain a contribution to the Raman spectrum similar to that of bulk graphite.

**Acknowledgments**

The authors thank G. Desiderio (CNR- IPCF LiCryL Laboratory) for his precious help in performing the SEM measurements. One of the authors (MG) acknowledges the financial support of the Regione Veneto within the Programma Operativo FSE 2006-13 of the European Union. Moreover, a financial support was granted by the Italian Ministry of Education, University, and Research to some of the authors at the University of Calabria within the program FIRB Futuro in Ricerca, project PLASMOGRAPH.

**Figure 1**

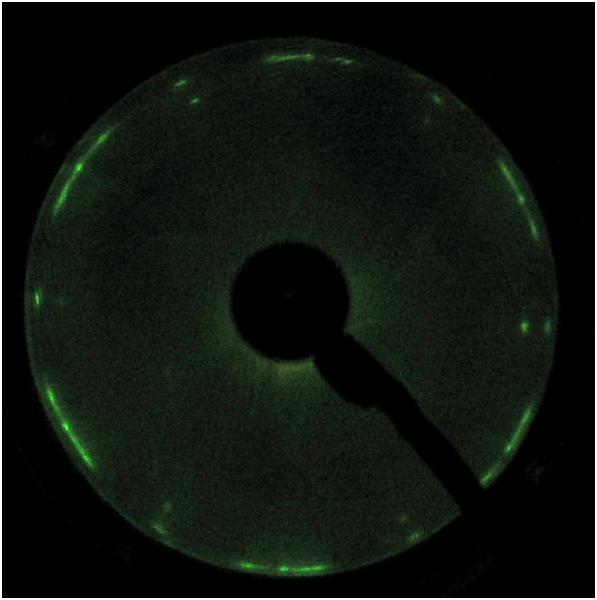

*Figure 1*: *LEED pattern of graphene on Pt(111), recorded at $E_p$ = 74.7 eV and for a sample temperature of 100 K.*

**Figure 2**

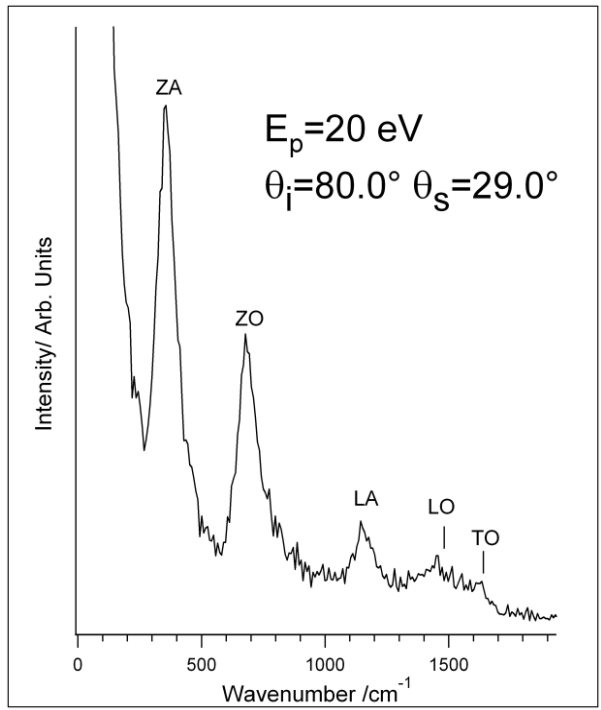

*Figure 2*: HREEL spectrum of the MLG/Pt(111) for an impinging energy of 20 eV. The incidence angle is 80.0° while the scattering angle is 29.0° (impact scattering conditions).

**Figure 3**

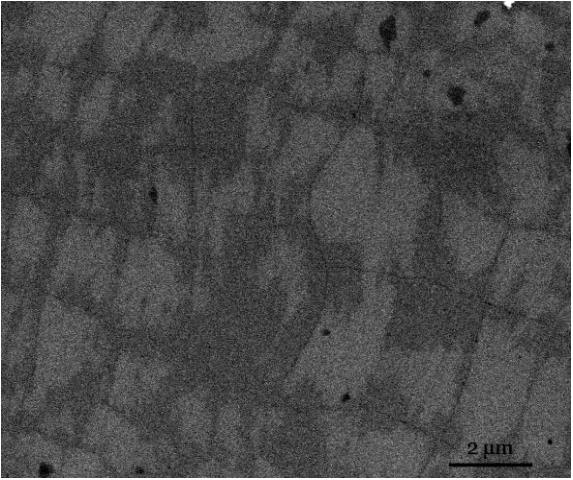

*Figure 3*: SEM image of a region of the graphene layer on a Pt(111) substrate. A wrinkle network is observed (darker lines) and two main different surface shades. Note also the few small black spots randomly distributed.

**Figure 4**

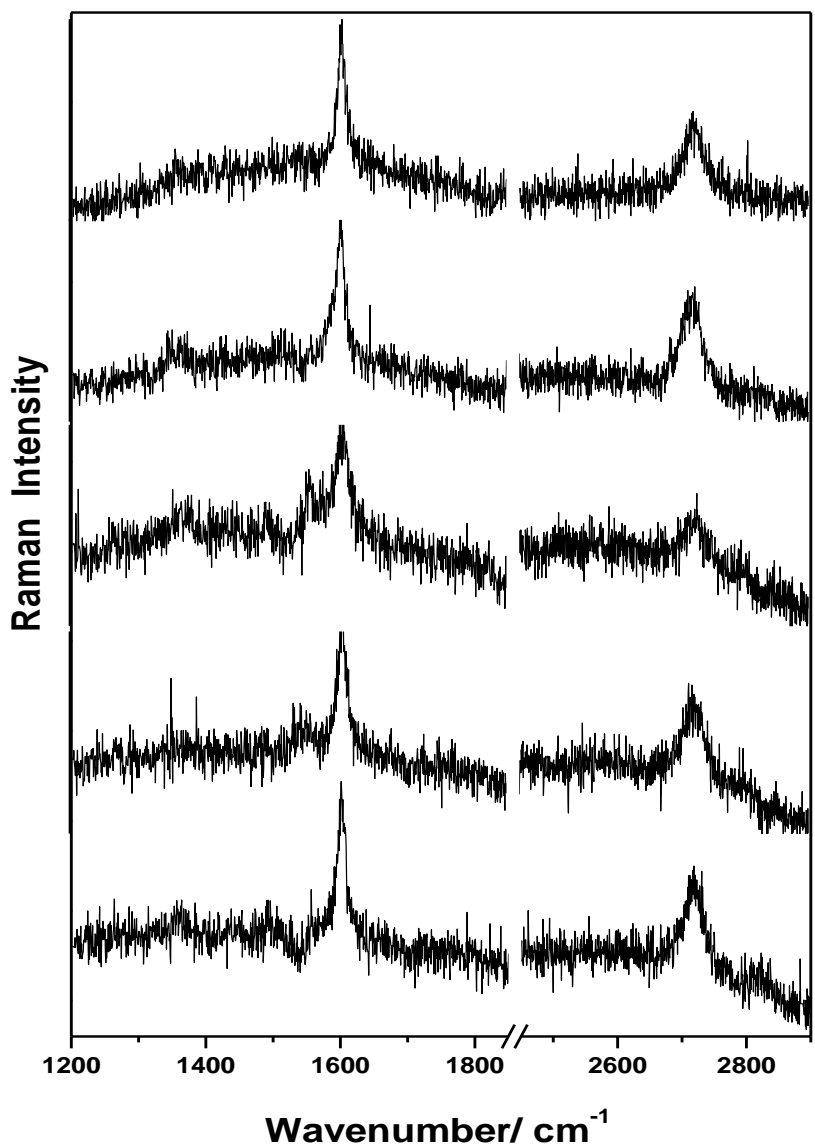

***Figure 4****: Raman spectra of graphene on Pt(111) recorded in crossed polarization (XY) from 5 different spots of the sample central region, along a right line with 10 μm spacing, under excitation wavelength at 488 nm, with a laser power of 20mw, and an integration time of 900 s. The spectral region including the D and G bands, and the one around the 2D overtone are plotted in the same intensity scale, while the intermediate range of no interest is not shown.*

**Figure 5**

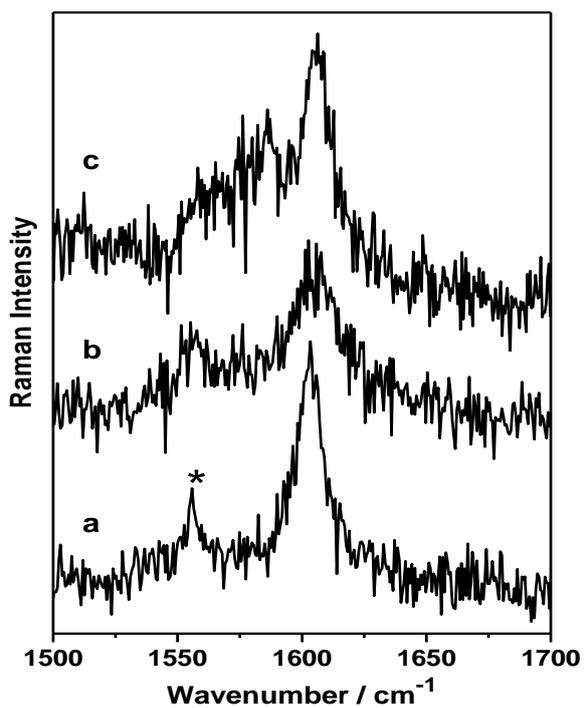

*Figure 5*: *Spectra collected from different spots, same conditions of laser power, integration time and polarization, represented with the same scale:*
*a) sample border, excitation at 488nm, asterisk indicate the $O_2$ stretching mode at 1556 $cm^{-1}$;*
*b) sample centre, excitation at 488nm;*
*c) sample centre, a different micro-region, excitation at 514.5 nm*